\documentclass{ws-procs9x6}

\def\be{\begin{equation}}
\def\ee{\end{equation}}

\def\bea{\begin{eqnarray}}
\def\eea{\end{eqnarray}}

\def\be{\begin{equation}}
\def\ee{\end{equation}}

\def\bea{\begin{eqnarray}}
\def\eea{\end{eqnarray}}

\def\gappeq{\mathrel{\rlap {\raise.5ex\hbox{$>$}}
{\lower.5ex\hbox{$\sim$}}}}
\def\lappeq{\mathrel{\rlap{\raise.5ex\hbox{$<$}}
{\lower.5ex\hbox{$\sim$}}}}

\def\ra{\rightarrow}

\def\mno{m_{\nu_1}}
\def\mnt{m_{\nu_2}}
\def\mnth{m_{\nu_3}}
\def\snu{\tilde{\nu}}

\def\lappeq{\mathrel{\rlap{\raise.5ex\hbox{$<$}}
{\lower.5ex\hbox{$\sim$}}}}

\def\mno{m_{\nu_1}}
\def\mnt{m_{\nu_2}}
\def\mnth{m_{\nu_3}}
\def\snu{\tilde{\nu}}

\def\CPbar{\hbox{{\rm CP}\hskip-1.20em{/}}~~}

\begin{document}
\title{Weak-scale implications of thermal leptogenesis in SUSY
\footnote{\uppercase{T}alk presented 
at {\it \uppercase{SUSY} 2003:
\uppercase{S}upersymmetry in the \uppercase{D}esert}\/, 
held at the \uppercase{U}niversity of \uppercase{A}rizona,
\uppercase{T}ucson, \uppercase{AZ}, \uppercase{J}une 5-10, 2003.
\uppercase{T}o appear in the \uppercase{P}roceedings. 
}}

\author{Sacha Davidson}

\address{Department of Physics \\
Durham University\\ 
Durham, DH1 3LE, GB\\ 
E-mail: sacha.davidson@durham.ac.uk}


\maketitle

\abstracts{Thermal leptogenesis is an attractive mechanism for
generating the baryon asymmetry of the Universe. However, 
in supersymmetric models, the parameter space is severely restricted
by the gravitino bound on the reheat temperature $T_{RH}$.
Using a parametrisation of the seesaw in terms of
left-handed inputs, which are related to
weak-scale observables in mSUGRA,  the low-energy footprints of
 thermal leptogenesis are discussed. 
}

\section{Introduction}

Neutrinos are observed to have small mass differences,
of order $10^{-5} - 10^{-2}$ eV$^{2}$.
Sadly, this is not  a prediction of 
Supersymmetry.  However, the seesaw\cite{seesaw}  mechanism is a natural
way to generate  such small majorana neutrino masses,
and, as an
added bonus, it provides ``for free'' a way to make 
the cosmological baryon asymmetry
(by leptogenesis\cite{FY,BP}).
The seesaw can easily be supersymmetrized.  

These proceeedings\footnote{ see
 \cite{DIP} for  more complete
references.}  summarise  (my) attempts to relate
leptogenesis to weak-scale observables. 
The approach is  ``bottom-up'': 
I want to avoid inputting a GUT/texture/
theoretical model for the structure of the Yukawa couplings
and mass matrices. 
They are based on work \cite{DI1,di2,phases,DIP,RK} with Alejandro
Ibarra and Ryuichiro Kitano, who I thank for illuminating
and productive collaborations. This proceedings
aims to be ``bedtime reading''; the paper\cite{DIP} is certainly longer and
I hope more careful.

Small neutrino masses and the observed baryon asymmetry of the Universe
(BAU) are evidence for Beyond-the-Standard-Model physics.  Can they
both be explained by the seesaw? And if yes, does
this have observable consequences?

The answer to the first question is ``yes''.
Unfortunately, there are no observable consequences of
generating the
BAU by leptogenesis in the Standard Model seesaw.
This is expected from parameter counting: the
high scale seesaw model has 18 parameters, 
whereas the effective light neutrino mass matrix 
has only 9. An  interesting approach, which has
been followed by many people, is therefore to
construct theoretically motivated models for 
the neutrino Yukawa matrix $Y_\nu$ and the majorana mass
matrix $ {\it M}$ of the 
heavy right-handed  neutrinos.

The aim here is different. We wanted to
study leptogenesis and the seesaw from a more
phenomenological ``bottom-up'' perspective---which is
possible in the supersymmetric (SUSY) version.  There are nine additional
low-energy parameters in the sneutrino mass matrix, which receives
contributions from  $Y_\nu$ and $ {\it M}$ through the Renomalisation
Group Equations (RGEs).
It is therefore possible {\it in principle} to reconstruct
the high scale seesaw parameters  $Y_\nu$ and $ {\it M}$
from the neutrino and sneutrino mass matrices. This ``reconstruction'' is
in practise impossible---it would require unrealistically accurate
measurements---but it is a useful {\it parametrisation} of
the seesaw. 

Using this parametrisation, we can express
the baryon asymmetry as a function of weak scale inputs, and
study the low-energy footprints of thermal leptogenesis.

\section{Notation and Assumptions}

The  leptonic superpotential in the seesaw can be written
\bea
\label{superp}
W = W_{MSSM}  
+ {\nu_R^c}^T {\bf Y_\nu} L\cdot H_u 
- \frac{1}{2}{\nu_R^c}^T{\it M}\nu_R^c , \eea
where $L_i$   are the left-handed 
lepton doublets, $H_u$ is the hypercharge $+1/2$
 Higgs doublet,  ${\bf Y_{\nu}}$ is the neutrino Yukawa matrix,
and  $\it M$ is a  Majorana mass matrix, with heavy eigenvalues
which are assumed hierarchical: $M_1 \ll M_2 < M_3$.

Two relevant bases for the $\nu_R$ vector space are the one where the
mass matrix ${ M}$ is diagonal ($ = { D_M}$), and where the Yukawa
matrix ${ Y_\nu Y_\nu^\dagger}$ is diagonal ($ = { D_Y^2}$). The unitary
matrix ${ V_R}$ transforms between these bases, so in the mass
eigenstate basis \be { Y_\nu Y_\nu^{\dagger}} = { V_R^\dagger D_Y^2
V_R}~~~.  \ee

At low energies, well below the $\nu_R$ mass scale, the light (LH)
neutrinos acquire an effective Majorana mass matrix $[m_\nu]$. In the
vector space of LH leptons, there are three interesting bases--- the one
where the charged lepton Yukawa $ { Y_e^\dagger Y_e} $ is diagonal, the
one where the neutrino Yukawa ${ Y_\nu^\dagger Y_\nu} $ is diagonal, and
the basis where $[m_\nu]$ is diagonal.  The first (${ D_{Y_e}}$) and last
(${ D_m}$) are phenomenologically important, and are related by the MNS
matrix ${ U} $:  $[m_\nu] = {U^* D_m U^\dagger} $ in the ${ D_{Y_e}}$ basis.
The second (${ D_{Y_\nu}}$) relative to the first (${ D_{Y_e}}$)
 can be important for phenomenology in SUSY
models, where ${ Y_\nu^\dagger Y_\nu}
(\equiv V_L^\dagger D_{Y_\nu}^2 V_L$ in the ${ D_{Y_e}}$ basis)
 induces flavour violation via its
appearance in the slepton RGEs.  The second(${ D_{Y_\nu}}$)
 and third(${ D_m}$) are
useful to relate LH and RH seesaw parameters\cite{DI1}, so
are appropriate for connecting weak-scale observables with
leptogenesis.

The matrix $W$ transforms between these bases. { 
In the  basis where ${ Y_\nu}$ is diagonal},  $[m_\nu]$ can be
written
\be
[m_\nu]  =  {D_Y M^{-1} D_Y}  v_u^2  = { W^* D_m W^\dagger } ~~~.
\label{kappa}
\ee 
The light neutrino masses are taken  hierarchical,
with $10^{-3} m_{\nu_2} < \mno < 0.1 \mnt$. It is assumed that
the largest eigenvalue of ${\bf Y_\nu}$, $y_3 \simeq 1$, and that
 there is   a steeper hierarchy in the eigenvalues of $Y_\nu$ than 
in those of the light neutrino mass matrix $[m_\nu]$.

Twenty-one parameters are required to fully determine the
Lagrangian of eqn (\ref{superp}). If $Y_e$ is neglected,
only 9 real numbers and 3 phases are required. These can
be chosen in various ways.
To relate the RH parameters relevant for leptogenesis to the
LH ones, many of which are accessible at low energy, it is useful to consider
the following possibilities:
\begin{enumerate}
\item `` top-down''---input the $\nu_R$ sector: ${ D_M}$, 
${ D_{Y_\nu Y_\nu^\dagger}},$ and ${ V_R}$.
\item `` bottom-up''---input the $\nu_L$ sector:  ${ D_\kappa}$, 
${ D_{Y_\nu^\dagger Y_\nu}},$ and ${ W}$.
\end{enumerate}

We assume gravity-mediated SUSY breaking, with universal soft 
masses at some scale $m_X \gg M_i$, and nothing but SUSY and the
seesaw between the electroweak scale and $m_X$ (so 
we know the RGEs). The leading log approximation for the slepton
mass matrix is used to relate angles of $V_L$ to
$\ell_j \rightarrow \ell_i \gamma $ branching ratios. 
So $W$ can be ``calculated''  from $\nu$ and $\tilde{\nu}$
mixing matrices.

\section{Leptogenesis}

The baryon asymmetry  produced via leptogenesis depends on the $\nu_R$
number density, the $\CPbar$ asymmetry in the $\nu_R$ decay, and
whether the decay is out of equilibrium. A cosmology-independent
way to produce the $\nu_R$
is by scattering in the thermal plasma after inflation. For hierarchical
right-handed neutrinos, this ``thermal leptogenesis'' scenario can be
described by 4 parameters \cite{BP}:  the lightest
$\nu_{R_1}$ mass $M_1$, its decay rate $\Gamma$
\footnote{the decay rate can be  rescaled to
be comparable to a light neutrino mass. The usual \cite{BP}
leptogenesis parameter is $\tilde{m}_1 = 8 \pi 
\Gamma \langle H_u^0 \rangle ^2/ M_1^2$.
}, which controls the
$\nu_{R_1}$ production and decay processes, the $\CPbar$ asymmetry
$\epsilon$ in the decay, and
an average neutrino mass $\bar{m}$ (which I do not discuss
here). There is an
upper bound on $\epsilon$ \cite{di2} (but see \cite{AS}):
\be
\epsilon = \frac{ \Gamma (\nu_R \rightarrow H \ell ) - 
\bar{\Gamma} (\nu_R \rightarrow \bar{H} \bar{\ell}) }{\Gamma + \bar{\Gamma}}
= \frac{8 \pi M_1 \mnth}{3 \langle H_u \rangle ^2} \delta ~~, \delta \leq 1 
\label{epsbd}
\ee

The BAU produced in thermal leptogenesis can  be written
\be
Y_B  = d( \Gamma) \epsilon  = \left\{ \begin{array}{ll}
3-9 \times 10^{-11} & BBN \\
7.5- 1.0 \times 10^{-11} & CMB \end{array} \right.
\ee
where $ d( \Gamma)$ is the ratio of the $\nu_R$ number density to the
entropy density, times  the fraction of the
produced lepton asymmetry which
survives as a baryon asymmetry today. 
 $ d(\Gamma)$ depends on the interactions of the $\nu_R$ in
the plasma, and has been numerically calculated \cite{BP}
to have a maximum value of $\sim 3 \times 10^{-4}$. 
A large enough BAU can be obtained if
\begin{equation}
\left( \frac{6 \times 10^{-11}}{Y_B} \right)
\left( \frac{d(g_*, \Gamma)}{3 \times 10^{-4} } \right)
\left( \frac{M_1}{10^9 GeV} \right)
\delta \mathrel{\rlap {\raise.5ex\hbox{$>$}}
{\lower.5ex\hbox{$\sim$}}} 1
\label{bd1}
\end{equation}

There are additional constraints on  the thermal leptogenesis
scenario in SUSY models. In gravity-mediated SUSY-breaking,
gravitino production imposes  an upper bound on the reheat temperature of
the Universe after inflation: $T_{RH} \lappeq 10^{9} - 10^{12}$
GeV. The canonical bound is
 $T_{RH} \lappeq 10^{9}$ GeV, and
\be
M_1 \lappeq T_{RH}
\label{bd2}
\ee
is required
to produce enough $\nu_R$.

\section{low-energy footprints}

In the parametrisation of  $Y_\nu$ and $ {\it M}$, in terms of
 $[m_\nu]$ and the sneutrino mass matrix  $[m_{\snu}^2]$, 
there is an analytic approximation for the the leptogenesis parameters
$M_1$,$ \Gamma$  and $\delta$, in terms of the light neutrino
masses $m_{\nu_i}$, a matrix $W =V_L U $ which rotates from the $
\nu_i$ mass eigenestate basis to the basis where $Y_\nu$ is diagonal
($\simeq$ rotation from the neutrino to sneutrino mass
eigenestate bases), and the smallest eigenvalue $y_1$ of
$Y_\nu$.

The low energy consequences of
thermal leptogenesis can be found by requiring eqns
(\ref{bd1}) and (\ref{bd2}) be satisfied.
This constrains $M_1$ to sit in
a narrow range around $ 10^{9}$ GeV, and $\epsilon$ to be
maximal.  $M_1 \sim 10^9$ GeV
determines $y_1$ as a function of $W$ and the
$m_{\nu_i}$. Since $y_1$ is effectively unmeasurable
in our parametrisation (it affects the first generation slepton masses
via the RGEs, which for $y_1 \sim 10^{-3} - 10^{-4} $
is a negligeable effect), this has no observable
consequences at low energy. The $\nu_R$ decay rate $\Gamma$
naturally  falls within the desirable range, so the low energy
consequences of  eqn (\ref{bd1}) correspond to  $\delta \ra 1$.

For $M_1 \sim 10^9$ GeV, $\delta$ must be
$O(1)$ (and $d_1$ maximal) to obtain a baryon asymmetry
at the lower end of the BBN range. This arises for $W$ near the
identity, which corresponds to mixing angles in the sleptons sector of order
the neutrino mixing angles.  This suggests that the
branching ratios for $\tau \ra \mu \gamma$ or 
 $\tau \ra e \gamma$ should be observable.  
From a model-building perspective, $W \sim {\it I}$ could
arise if the large MNS angles arise from diagonalising
the charged lepton Yukawa $Y_e$. 

For $M_1 \sim 10^{10}$ GeV, a large enough baryon asymmetry can be
obtained for $W \sim U$,  provided that   $W_{13} \sim 0.04$.
This corresponds to
 an observable CHOOZ angle $\theta_{13} \sim 0.04$,
or observable  $\tau \ra e \gamma$, ...or to no
observable consequences at all (It is unfortunately
possible to have 
$W_{13} \sim 0.04$ with  arbitrarily small CHOOZ angle and
lepton flavour violating branching ratios).  The case
$W \sim U$  arises in many  models,  where the large mixing
angles of the neutrino sector come from diagonalising
$[m_\nu]$ in the ``texture'' basis.

\begin{figure}
\psfig{figure=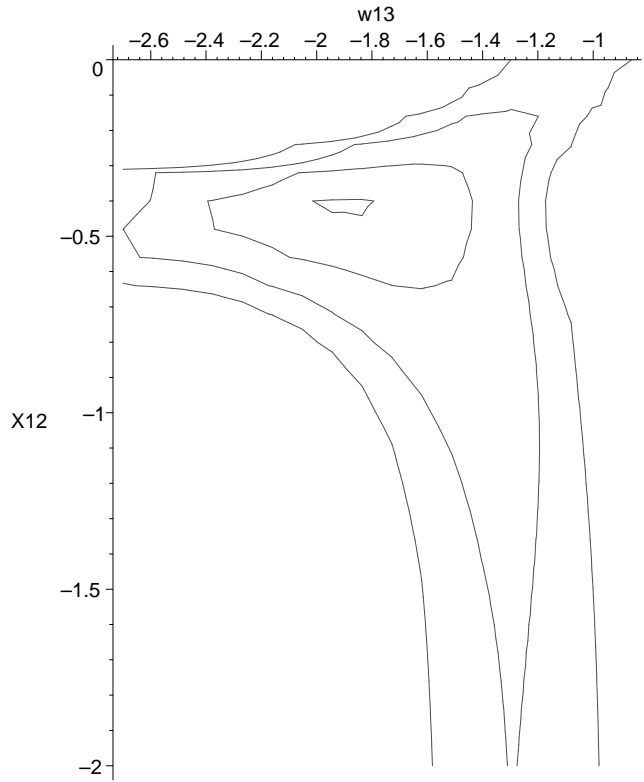,height=4in}
\caption{
Contour plot  of $Y_B$, 
as a function of $\omega_{13} \simeq \log [W_{13}]$,  and 
$\chi_{12} \simeq \log [ V_{L12} + V_{L13}]$.
 The contours enclose the area when 
$Y_B > 2 \times 10^{-11}$, for $M_1 =f \times  10^9 $ GeV,
central values of $m_{\nu_3}$ and $m_{\nu_2}$,  
and $m_{\nu_1} = m_{\nu_2}/10$. In
the direction of increasing area, the lines correpond
to $f = 1, 3, 6$ and 9.
\label{radish}}
\end{figure}

Figure \ref{radish} shows contours of constaint $Y_B$, labelled
by $f = 1,3,6$ and 9. $Y_B \gappeq 2 \times 10^{-11}$ inside
the curve, for $M_1 = f \times 10^{9}$ GeV. The variables 
on the axes are chosen to provide as ``physical'' a measure
on parameter space as possible. They are vaguely related
to logarithms of measurable quantities: 
 $\omega_{13}  \sim \theta_{13} + 
\sqrt{10^6 BR( \tau \ra e \gamma)} + $ something
unmeasurable, and 
$\chi_{12}  
 \sim \sqrt{10^6 BR( \tau \ra e \gamma)}
+ \sqrt{10^6 BR( \tau \ra \mu \gamma)}$.

\section{Summary}

Thermal leptogenesis can work in supersymmetric seesaw models.
It makes low energy predictions because the available parameter
space is restricted. Observing lepton flavour violating
decays, such as $\tau \ra \ell \gamma$, or
a CHOOZ angle $ \sim 0.04$ would
lend support to this scenario.

\section*{Acknowledgments}
I am grateful to the organisers  
for organising this memorable member of the SUSY series,
and for the invitation. 
I also thank many participants for interesting discussions.

This work was supported in part by a PPARC Advanced
Fellowship.


\end{document}